\documentclass[aps,prd,superscriptaddress,amsmath,groupedaddress,       
         twocolumn,showpacs]{revtex4}

\usepackage{amssymb,graphicx}
\usepackage{dcolumn}
\usepackage{rotating}

\newcommand{\msb}{{\rm\overline{MS}}}
\newcommand{\alphamsb}{\alpha_\msb}
\newcommand{\eq}[1]{Eq.~(\ref{#1})}
\newcommand{\order}{{\cal O}}

\newcommand{\alphav}{\alpha_V}
\newcommand{\qcd}{\mathrm{QCD}}
\newcommand{\tr}{\mathrm{Tr}}
\newcommand{\sigterm}[1]{c_{#1}\,\sigma^2_{#1}}

\newcommand{\almz}{0.1183\,(8)}
\newcommand{\alv}{0.2120\,(28)}
\newcommand{\chialmz}{0.2}

\begin{document}

\title{Update: Accurate Determinations of $\alpha_s$ from Realistic 
Lattice QCD}
\author{C.\ T.\ H.\ Davies}
\affiliation{Department of Physics and Astronomy, University of 
Glasgow, Glasgow G12 8QQ, UK}
\author{K.\ Hornbostel}
\affiliation{Southern Methodist University, Dallas, Texas 75275, USA}
\author{I.\ D.\ Kendall}
\affiliation{Department of Physics and Astronomy, University of 
Glasgow, Glasgow G12 8QQ, UK}
\author{G.\ P.\ Lepage}
\email{g.p.lepage@cornell.edu}
\affiliation{Laboratory for Elementary-Particle Physics, Cornell 
University, Ithaca, NY 14853, USA}
\author{C.\ McNeile}
\affiliation{Department of Physics and Astronomy, University of 
Glasgow, Glasgow G12 8QQ, UK}
\author{J.\ Shigemitsu}
\affiliation{Physics Department, The Ohio State University, Columbus, 
Ohio 43210, USA}
\author{H.\ Trottier}
\affiliation{Physics Department, Simon Fraser University, Vancouver, 
British Columbia, Canada}
\collaboration{HPQCD Collaboration}
\date{17 August 2008}

\begin{abstract}
We use lattice QCD simulations, with MILC configurations (including 
vacuum polarization from $u$, $d$, and $s$ quarks), to update our 
previous determinations of the QCD coupling constant. Our new analysis 
uses results from 6~different lattice spacings and 12~different 
combinations of sea-quark masses to significantly reduce our previous 
errors. We also correct for finite-lattice-spacing errors in the scale 
setting, and for nonperturbative chiral corrections to the 
22~short-distance quantities from which we extract the coupling. Our 
final result is $\alphav(7.5\,\mathrm{GeV},n_f\!=\!3) = \alv$, which is 
equivalent to $\alphamsb(M_Z,n_f\!=\!5)= \almz$. We compare this with 
our previous result from Wilson loops, which differs by one standard 
deviation.
\end{abstract}

\maketitle

\section{Introduction} \label{sec:introduction}
An accurate value for the coupling constant $\alpha_s$ in quantum 
chromodynamics (QCD) is important both for QCD phenomenology, and as an 
input for possible theories beyond the Standard Model. Some of the most 
accurate values for the coupling constant come from numerical 
simulations of QCD using lattice techniques, when combined with very 
accurate experimental data for hadron masses. In this paper we update 
our previous determinations of the coupling from Wilson loops in 
lattice QCD~\cite{Mason:2005zx}. Our new analysis takes advantage of 
new simulation results, from the MILC collaboration, that employ 
smaller lattice spacings~$a$. We also now account systematically for 
chiral corrections associated with the masses of sea quarks in the 
simulation, and for $\order(a^n)$ uncertainties in the values we use 
for the lattice spacing.

Few-percent accurate QCD simulations have only become possible in the 
last few years, with the development of much more efficient techniques 
for simulating the sea quarks; see, for example,~\cite{Davies:2003ik} 
for an overview and references. The simulations we use include only 
light quarks ($u$, $d$ and $s$) in the vacuum polarization; the effects 
of $c$ and $b$ quarks are incorporated using perturbation theory, which 
is possible because of their large masses. Our lattice QCD analysis 
proceeds in two steps. First the QCD parameters\,---\,the bare coupling 
constant and bare quark masses in the Lagrangian\,---\,must be tuned. 
For each value of the bare coupling, we set the lattice spacing to 
reproduce the correct $\Upsilon^\prime$--$\Upsilon$~meson mass 
difference in the simulations, while we tune the $u/d$, $s$, $c$ 
and~$b$ masses to give correct values for $m_\pi^2$, $2m^2_K-m_\pi^2$, 
$m_{\eta_c}$, and $m_\Upsilon$, respectively; more information can be 
found in \cite{Davies:2003ik}. For efficiency we set $m_u=m_d$; this 
leads to negligible errors in the analysis presented here. Once these 
parameters are set, there are no further physics parameters, and the 
simulation will accurately reproduce QCD.

Having an accurately tuned simulation of QCD, we use it to compute 
nonperturbative values for a variety of short-distance quantities, each 
of which has a perturbative expansion of the form
\begin{equation}\label{Y-pth}
	Y = \sum_{n=1}^\infty c_n\alpha_V^n(d/a)
\end{equation}
where $c_n$ and $d$ are dimensionless $a$-independent constants, and 
$\alphav(d/a)$ is the (running) QCD coupling constant, with $n_f=3$ 
light-quark flavors, in the 
$V$~scheme~\cite{Lepage:1992xa,Brodsky:1982gc}. Given the coefficients 
$c_n$, which are computed using Feynman diagrams, we choose 
$\alphav(d/a)$ so that the perturbative formula for $Y$ reproduces the 
nonperturbative value given by the simulation. Given $d$ and $a$, and 
the $c$ and $b$ masses, we can then use perturbation theory to convert 
$\alphav(d/a)$ to the more conventional coupling constant 
$\alphamsb(M_Z,n_f\!=\!5)$, evaluated at the mass of the 
Z~meson~\cite{heavy-quarks,Schroder:1998vy}.

This analysis is complicated by nonperturbative contributions to $Y$ 
and by simulation uncertainties in the value of the lattice 
spacing~$a$, which enters~\eq{Y-pth}. It is also complicated by 
perturbative uncertainties. We know the values of the coefficients 
$c_n$ through order $n=3$ (next-to-next-to-leading order) for the 
quantities we examine, yet unknown higher-order coefficients still have 
an impact at the level of accuracy we seek. A main focus of this paper 
is to address these complications, and quantify the uncertainties in 
our determination of the coupling constant. In 
Section~\ref{sec:perturbation_theory} we review the perturbative 
expansions for our short-distance quantities, all but one of which are 
derived from small Wilson loops~\cite{no-V}. The Monte Carlo simulation 
results for these loops are presented in 
Section~\ref{sec:qcd_simulations}.
We discuss finite-lattice-spacing errors and chiral corrections in 
Section~\ref{sec:systematic_errors}. In 
Section~\ref{sec:analysis_and_results}, we describe how we combine 
perturbation theory with simulation results using constrained 
(Bayesian) fitting methods. There we present our results and discuss in 
detail the various uncertainties that arise. Finally, in 
Section~\ref{sec:conclusions}, we summarize our results.

\section{Perturbation Theory} \label{sec:perturbation_theory}
The simplest short-distance quantities to simulate are vacuum 
expectation values of Wilson loop operators:
\begin{equation}\label{eq:wilson-loop}
W_{mn} \equiv \mbox{$\frac{1}{3}$}\,\langle0|\, 
\mathrm{Re\,Tr}\,\mathrm{P}\,\mathrm{
e}^{-ig\oint_{nm}
\!A\cdot dx}\, |0\rangle,
\end{equation}
where ${\rm P}$~denotes path ordering, $A_\mu$~is the QCD vector
potential, and the integral is over a closed
${ma\!\times\!na}$~rectangular path. Wilson loops should be calculable 
in (lattice QCD) perturbation theory when~$ma$ and~$na$ are small. We 
computed perturbative coefficients through order $n=3$ for six small, 
rectangular loops, and also for two non-planar paths:
\vspace{-2ex}
\begin{equation}\label{bent-paths}
\mathrm{BR} =
\begin{picture}(60,30)(0,15)
  \put(10,10){\vector(0,1){12.5}}
  \put(10,10){\line(0,1){20}}
  \put(10,30){\vector(2,1){10}}
  \put(10,30){\line(2,1){15}}
  \put(25.2,37.6){\vector(1,0){12.5}}
  \put(25.2,37.6){\line(1,0){20}}
  \put(45.2,37.6){\vector(0,-1){12.5}}
  \put(45.2,37.6){\line(0,-1){20}}
  \put(45.2,17.6){\vector(-1,0){12.5}}
  \put(45.2,17.6){\line(-1,0){20}}
  \put(25.2,17.6){\vector(-2,-1){10}}
  \put(25.2,17.6){\line(-2,-1){15}}
\end{picture}
\quad
\mathrm{CC} =
\begin{picture}(60,30)(0,15)
  \put(10,10){\vector(0,1){12.5}}
  \put(10,10){\line(0,1){20}}
  \put(10,30){\vector(2,1){10}}
  \put(10,30){\line(2,1){15}}
  \put(25.2,37.6){\vector(1,0){12.5}}
  \put(25.2,37.6){\line(1,0){20}}
  \put(45.2,37.6){\vector(0,-1){12.5}}
  \put(45.2,37.6){\line(0,-1){20}}
  \put(45.2,17.6){\vector(-2,-1){10}}
  \put(45.2,17.6){\line(-2,-1){15}}
  \put(30,10){\vector(-1,0){12.5}}
  \put(30,10){\line(-1,0){20}}
\end{picture}.
\end{equation}
The coefficients for our various loops are derived 
in~\cite{loop-paper}. The results are for the gluon and quark actions 
used to create the MILC gluon-configuration sets used in this study. 
They also assume $n_f=3$ massless sea quarks. The quarks in our 
simulations are not exactly massless, but the masses are sufficiently 
small that the difference is negligible, $\order(\alphav^2 (am)^2)$, in 
perturbation theory (but less so nonperturbatively, as we will 
discuss).

\begin{table*}
	\caption{Perturbative scale and coefficients for several small Wilson 
loops~$W_{ij}$, Creutz
	ratios, tadpole-improved Wilson loops,
	and the tadpole-improved bare coupling~$\alpha_\mathrm{lat}/W_{11}$.
	Parameters $d$ and $c_i$ are defined in~\eq{Y-pth}. Coefficients 
$c_1,c_2,c_3$ are
	from lattice perturbation theory; coefficients $c_4,c_5$ are from
	the fits to results from multiple lattice spacings described in this 
paper.
	These results are for the $a^2$-improved
	gluon action used by the MILC collaboration,
	with the ASQTAD action for vacuum polarization from $n_f=3$ massless 
quarks. Similar types of
	short-distance quantity are grouped.}
	\label{tab:pth-coef}
	\begin{ruledtabular}
		\begin{tabular}{rrddd|ll}
		& \multicolumn{1}{c}{$d$}
		& \multicolumn{1}{c}{\quad$c_1$}
		& \multicolumn{1}{c}{\quad\quad$c_2/c_1$}
		& \multicolumn{1}{c}{\quad$c_3/c_1$}
		& \multicolumn{1}{l}{\:\:\:$c_4/c_1$}
		& \multicolumn{1}{l}{\:\:\:$c_5/c_1$} \\ \hline
		$-\log W_\mathrm{11}$ &  3.325 &    3.06840 &     -1.0683\,(2) &   
1.70\,(4) & $-4$\,(2) & $-0$\,(4) \\
$-\log W_\mathrm{12}$ &  2.998 &    5.55120 &     -0.8585\,(4) &   
1.72\,(4) & $-4$\,(2) & $-1$\,(4)\\
$-\log W_\mathrm{BR}$ &  3.221 &    4.83425 &     -0.8547\,(3) &   
1.80\,(4)  & $-4$\,(2) & $-1$\,(4)\\
$-\log W_\mathrm{CC}$ &  3.047 &    5.29758 &     -0.7941\,(3) &   
1.86\,(4)  & $-4$\,(2) & $-1$\,(5)\\
$-\log W_\mathrm{13}$ &  2.934 &    7.87656 &     -0.7437\,(8) &   
1.75\,(5)  & $-4$\,(2) & $-1$\,(4)\\
$-\log W_\mathrm{14}$ &  2.895 &   10.17158 &     -0.6870\,(8) &   
1.70\,(6)  & $-4$\,(2) & $-1$\,(4)\\
$-\log W_\mathrm{22}$ &  2.582 &    9.19970 &    -0.6923\,(10) &   
1.86\,(5)  & $-4$\,(2) & $-1$\,(4)\\
$-\log W_\mathrm{23}$ &  2.481 &   12.34282 &    -0.5995\,(13) &   
2.00\,(6)  & $-4$\,(2) & $-1$\,(5)\\ \\
$-\log W_{13}/W_{22}$ &  2.397 &   -1.32313 &     0.5969\,(84) &  
1.11\,(21) & $-2$\,(2) & $-1$\,(3) \\
$-\log W_{11}W_{22}/W_{12}^2$ &  2.169 &    1.16569 &     0.7361\,(86) 
&  1.21\,(22) & $-2$\,(2) & $-1$\,(3)\\
$-\log W_\mathrm{CC}W_\mathrm{BR}/W_{11}^3$ &  2.728 &    0.92665 &     
2.2825\,(19) &   0.78\,(9) & $-4$\,(4) & $-2$\,(6)	\\
$-\log W_\mathrm{CC}/W_\mathrm{BR}$ &  2.730 &    0.46333 &     
0.5103\,(35) &  1.16\,(12) & $-2$\,(2)  & $-1$\,(3)\\
$-\log W_\mathrm{14}/W_\mathrm{23}$ &  2.066 &   -2.17124 &     
0.5838\,(84) &  1.83\,(29) & $-3$\,(3)  & $-1$\,(4)\\
$-\log W_{11}W_{23}/W_{12}W_{13}$ &  1.970 &    1.98345 &     
0.7062\,(88) &  1.64\,(27)   & $-3$\,(3)  & $-1$\,(4)\\ \\
$-\log W_\mathrm{12}/u_0^{6}$ &  2.470 &    0.94861 &     0.6011\,(19) 
&   0.05\,(8)       & $-3$\,(2)  & $-1$\,(2)\\
$-\log W_\mathrm{BR}/u_0^{6}$ &  2.720 &    0.23166 &     4.0516\,(41) 
&  0.36\,(16)       & $-8$\,(6) & $-3$\,(10)\\
$-\log W_\mathrm{CC}/u_0^{6}$ &  2.730 &    0.69499 &     1.6925\,(20) 
&   0.91\,(8)       & $-3$\,(3)  & $-1$\,(4)\\
$-\log W_\mathrm{13}/u_0^{8}$ &  1.888 &    1.73977 &     0.4019\,(34) 
& -0.44\,(10)       & $-2$\,(1)  & $-1$\,(2)\\
$-\log W_\mathrm{14}/u_0^{10}$ &  1.892 &    2.50059 &     0.4817\,(33) 
& -0.68\,(15)      & $-2$\,(1)  & $-1$\,(2)\\
$-\log W_\mathrm{22}/u_0^{8}$ &  2.290 &    3.06291 &     0.6149\,(30) 
&   0.44\,(9)       & $-2$\,(2)  & $-1$\,(2)\\
$-\log W_\mathrm{23}/u_0^{10}$ &  2.030 &    4.67183 &     0.5714\,(35) 
&  0.55\,(11)      & $-2$\,(2)  & $-1$\,(2)\\ \\
$\alpha_\mathrm{lat}/W_{11}$ &  3.325 &    1.00000 &     -0.4212\,(2) & 
  0.72\,(4)        & $-4$\,(1)  & $-1$\,(2)\\

	\end{tabular}
	\end{ruledtabular}
\end{table*}

Perturbation theory is more convergent for the logarithm of a Wilson 
loop than it is for the loop itself. This is because the perturbative 
expansion of a loop is dominated by a self-energy contribution that is 
proportional to the length of the loop, and this contribution 
exponentiates for large loops. The length of the loop factors out of 
the expansion when we take the logarithm. This structure is evident in 
Table~\ref{tab:pth-coef} where we tabulate the perturbative 
coefficients for the logarithms of our loops. The renormalization 
scales~$d/a$ for each quantity are determined using the procedures 
described in~\cite{Lepage:1992xa,Brodsky:1982gc,Hornbostel:2002af}.

The perturbative coefficients in $\log(W)$, while greatly reduced by 
the logarithm, are still rather large. They can be further reduced in 
two ways. One is to ``tadpole improve'' $W_{mn}$ by dividing by 
$u_0^{2(n+m)}$ where~\cite{Lepage:1992xa}
\begin{equation}\label{u0-def}
	u_0\equiv(W_{11})^{1/4}.
\end{equation}
The other is to examine Creutz ratios of the loops rather than the 
loops themselves~\cite{Lepage:1992xa}. Each procedure significantly 
reduces the known high-order coefficients, as is clear in 
Table~\ref{tab:pth-coef}.
We use seven tadpole-improved loops and six Creutz ratios in our 
analysis. Each has smaller $\alphav^3$ coefficients, which improves 
convergence, but each also has a significantly smaller scale~$d/a$, 
which slows convergence (since $\alphav(d/a)$ is larger).

We also include in Table~\ref{tab:pth-coef} the perturbative expansion 
for the tadpole-improved bare coupling constant, 
$\alpha_\mathrm{lat}/W_{11}$, where $\alpha_\mathrm{lat}$ is the 
coupling constant that appears in the gluon action for a given lattice 
spacing~\cite{Lepage:1992xa}. This is another, independent, 
short-distance quantity from which $\alphav$ can be determined.

We used Feynman diagrams to compute perturbative coefficients $c_n$ for 
$n\le3$. Higher-order coefficients can be estimated by simultaneously 
fitting results from different lattice spacings to the same 
perturbative formula~\cite{Mason:2005zx}. This is possible because the 
coupling $\alphav(d/a)$ changes value with different lattice 
spacings~$a$:
\begin{equation}\label{evol-eq}
	q^2\,\frac{d\alphav(q)}{dq^2} = - \beta_0 \alphav^2 - \beta_1 
\alphav^3
	-\beta_2 \alphav^4 -\beta_3\alphav^5
\end{equation}
where the $\beta_i$ are constants~\cite{Schroder:1998vy}. In this 
paper, we follow our earlier analysis by parameterizing the running 
coupling by its value at 7.5\,GeV,
\begin{equation}
	\alpha_0 \equiv \alphav(7.5\,\mathrm{GeV},n_f\!=\!3).
\end{equation}
Given $\alpha_0$, the coupling at any other scale can be obtained by 
integrating \eq{evol-eq} (which we do numerically).

For the purposes of this paper, we define $\alphav$ in fourth order and 
beyond so that the evolution equation, \eq{evol-eq}, is exact, with no 
higher-order terms beyond~$\beta_3$. This definition gives precise 
meaning to the perturbative coefficients $c_n$ for $n\ge4$ that we 
determine by fitting the $a$-dependence of our short-distance 
quantities~\cite{lepage:erratum}.

Our main result is a value for $\alpha_0$. To facilitate comparisons 
with other analyses, we convert this result to the 
$\msb$~scheme~\cite{Schroder:1998vy}, add in $c$ and $b$ vacuum 
polarization perturbatively~\cite{heavy-quarks}, and then evolve to the 
mass of the $Z$~meson, again using perturbation 
theory~\cite{Schroder:1998vy}.

\section{QCD Simulations} \label{sec:qcd_simulations}
The gluon-configuration sets we use were created by the MILC 
collaboration~\cite{MILC}. The relevant simulation parameters are 
listed in Table~\ref{tab:qcd-param}.

The input parameters for a QCD simulation are the bare coupling 
constant and bare quark masses. The coupling constant is specified 
through the $\beta$~parameter, listed in Table~\ref{tab:qcd-param}, 
where
\begin{equation}
	\alpha_\mathrm{lat} \equiv \frac{5}{2\pi\beta}.
\end{equation}
The bare quark masses, $m_{0\ell}(a)$ for $u/d$~quarks and $m_{0s}(a)$ 
for $s$~quarks, used in the simulations are also listed, in units of 
the lattice spacing and, following MILC conventions, multiplied 
by~$u_0$~(\eq{u0-def}). The bare masses corresponding to fixed physical 
masses (of, for example, pions) vary with the lattice spacing. To 
facilitate comparisons between lattice spacings, we use first-order 
perturbation theory to evolve all of our masses to a common value for 
the lattice spacing, which we take to be the smallest lattice spacing 
in our analysis:
\begin{equation}
	m_q \equiv m_{0q}(a_\mathrm{min})
\end{equation}
The $s$-quark masses here are approximately correct. The $u/d$~masses 
are generally too large, but small enough to allow accurate 
extrapolations to the correct values.

The lattice spacing is not an input to QCD simulations. Rather it is 
extracted from calculations of physical quantities in the simulation. 
Here we use MILC's determinations of $r_1/a$ for this purpose, where 
$r_1$ is defined in terms of the static-quark potential~\cite{MILC}. 
The values for each configuration set are listed in 
Table~\ref{tab:qcd-param}. To obtain the lattice spacing, we need to 
know~$r_1$. We use the value, $r_1=0.321\,(5)$\,fm, determined from 
simulation results for the $\Upsilon^\prime$--$\Upsilon$ mass 
splitting~\cite{Gray:2005ur}. The uncertainties quoted for $r_1/a$ in 
Table~\ref{tab:qcd-param} are predominantly statistical; they {do not} 
include potential errors due to the finite lattice spacing or mistuned 
light-quark masses, which we will discuss later.

The lattices we use here have lattice spacings that range from 0.18\,fm 
to~0.045\,fm. The spatial volumes are 2.4\,fm across or larger in each 
case.

\begin{table}
	\squeezetable
	\caption{QCD parameters for the 12~different sets of gluon 
configurations used in this paper~\cite{MILC}. Parameter $\beta$ 
specifies the bare coupling constant. The inverse lattice spacing is 
specified in terms of the $r_1$, and the bare quark masses are in units 
of the lattice spacing, and multiplied by~$u_0$. The spatial and 
temporal sizes, $L$ and $T$, are also given. Configuration sets that 
were tuned to have the same lattice spacing are grouped.}
	\label{tab:qcd-param}
	\begin{ruledtabular}
		\begin{tabular}{rllllrr}
			Set & \multicolumn{1}{c}{$\beta$}
			& \multicolumn{1}{c}{$r_1/a$}
			& \multicolumn{1}{c}{$au_0m_{0\ell}$}
			& \multicolumn{1}{c}{$au_0m_{0s}$}
			& \multicolumn{1}{c}{$L/a$}
			& \multicolumn{1}{c}{$T/a$}
			\\ \hline
			1 & 6.458 &  1.802(10)  & 0.0082  & 0.082   &   16  & 48 \\ \\
			2 & 6.572 &  2.133(14)  & 0.0097  & 0.0484  &  16   & 48 \\
			3 & 6.586 &  2.129(12)  & 0.0194  & 0.0484  &  16   & 48 \\ \\
			4 & 6.76  &  2.632(13)  & 0.005   & 0.05    &  24   & 64 \\
			5 & 6.76  &  2.610(12)  & 0.01    & 0.05    &  20   & 64 \\
			6 & 6.79  &  2.650(08)  & 0.02    & 0.05    &  20   & 64 \\ \\
			7 & 7.09  &  3.684(12)  & 0.0062  & 0.031   &  28   &  96 \\
			8 & 7.11  &  3.711(13)  & 0.0124  & 0.031   &  28   &  96 \\ \\
			9 & 7.46  &  5.264(13)  & 0.0018  & 0.018   &  64   & 144 \\
			10 & 7.47  &  5.277(16)  & 0.0036  & 0.018   &  48   & 144 \\
			11 & 7.48  &  5.262(22)  & 0.0072  & 0.018   &  48   & 144 \\ \\
			12 & 7.81  &  7.127(34)  & 0.0028  & 0.014   &  64   & 192
		\end{tabular}
	\end{ruledtabular}
\end{table}

Our simulation results for the vacuum expectations of our 8~different 
Wilson loops, each for each of our 12~different configuration sets, are 
presented in Table~\ref{tab:w-mc}. The uncertainties quoted are 
statistical. Step-size errors, due to the algorithm used to generate 
gluon configurations, are no larger than the statistical 
errors~\cite{Aubin:2004wf} and therefore, like statistical errors, are 
negligible; we will ignore them here.

\begin{table*}
	\squeezetable
	\caption{Simulation results for the vacuum expectation values of 
various small Wilson loops. Results are given for each of the 
12~different configuration sets in Table~\ref{tab:qcd-param}.}
	\label{tab:w-mc}
	\begin{ruledtabular}\begin{tabular}{rllllllll}
	Set	   & $W_\mathrm{11}$ & $W_\mathrm{12}$ & $W_\mathrm{13}$ & 
$W_\mathrm{14}$ & $W_\mathrm{22}$ & $W_\mathrm{23}$ & $W_\mathrm{BR}$ & 
$W_\mathrm{CC}$ \\ \hline
		 1  & 0.534101(17) & 0.280720(22) & 0.149263(21) & 0.079710(19) & 
0.087438(23) & 0.030150(16) & 0.338982(22) & 0.287376(25) \\
		 2  & 0.548012(51) & 0.298624(68) & 0.165063(67) & 0.091701(63) & 
0.101572(73) & 0.038333(54) & 0.356763(68) & 0.306315(78) \\
		 3  & 0.549470(53) & 0.300310(70) & 0.166530(70) & 0.092797(63) & 
0.102640(76) & 0.039007(54) & 0.358570(70) & 0.308140(79) \\
		 4  & 0.567069(16) & 0.323163(21) & 0.187281(24) & 0.109122(27) & 
0.121542(19) & 0.050751(15) & 0.381148(25) & 0.332184(27) \\
		 5  & 0.566961(21) & 0.322987(27) & 0.187084(26) & 0.108927(21) & 
0.121341(29) & 0.050579(21) & 0.380988(26) & 0.331996(29) \\
		 6  & 0.569716(21) & 0.326496(27) & 0.190278(26) & 0.111479(21) & 
0.124204(29) & 0.052397(21) & 0.384479(26) & 0.335685(29) \\
		 7  & 0.594843(7) & 0.359761(9) & 0.221624(10) & 0.137271(10) & 
0.153433(12) & 0.072261(10) & 0.417002(9) & 0.370239(10) \\
		 8  & 0.596408(12) & 0.361838(19) & 0.223616(17) & 0.138946(16) & 
0.155315(18) & 0.073593(16) & 0.419020(16) & 0.372372(17) \\
		 9  & 0.620813(5) & 0.394837(8) & 0.255897(9) & 0.166723(9) & 
0.186116(7) & 0.096208(6) & 0.450947(7) & 0.406300(8) \\
		10  & 0.621462(3) & 0.395717(4) & 0.256770(5) & 0.167486(5) & 
0.186959(5) & 0.096852(5) & 0.451798(4) & 0.407210(5) \\
		11  & 0.622115(2) & 0.396607(4) & 0.257650(4) & 0.168257(4) & 
0.187809(5) & 0.097491(4) & 0.452654(3) & 0.408123(4) \\
		12  & 0.641947(2) & 0.423992(3) & 0.285304(5) & 0.192943(5) & 
0.214759(4) & 0.118532(3) & 0.478903(3) & 0.436064(4) \\
	\end{tabular}\end{ruledtabular}
\end{table*}

\section{Systematic Errors} \label{sec:systematic_errors}
The goal of our analysis is to determine 
$\alpha_0\equiv\alphav(7.5\,\mathrm{GeV})$.
The only relevant systematic errors, other than from the truncation of 
perturbation theory, are  from nonperturbative effects and from $a^2$ 
errors in our determination of the lattice spacings. Finite-volume 
errors are no larger than our statistical errors, as we have verified 
by examining configuration set~5 with $L/a=28$ in addition to $L/a=20$. 
Statistical errors are also negligible (and therefore we ignored 
statistical correlations between different Wilson loops when computing 
Creutz ratios, whose real statistical errors are 2--3~times smaller 
than what we use here). We consider each systematic effect in term.

	\subsection{Chiral Corrections} 	\label{sub:chiral_corrections}
	Wilson loops, being very short-distance, are almost independent of the 
light-quark masses. The dependence in perturbation theory is 
$\order(\alphav^2(am_q)^2)$, which is negligible here given other 
errors. There is a larger contribution, however, from nonperturbative 
contributions that is important to our analysis. This contribution can 
be parameterized using chiral perturbation theory and the operator 
product expansion, which says that an arbitrary QCD operator $O_\qcd$ 
that is local at scale~$\Lambda$ can be expanded in terms of local 
operators $O_n$ from the chiral theory:
\begin{equation}
	O_\qcd \equiv \sum_n b_n \frac{O_n}{\Lambda^{d_n}}
\end{equation}
where $d_n$ is the dimension of $O_n$ minus the dimension of $O_\qcd$. 
Here equivalence between the left-hand and right-hand sides means that 
matrix elements of the operators are equal for comparable physical 
states in QCD and the chiral theory.

	For Wilson loops, we are interested in vacuum expectation values and 
singlet operators. The scale $\Lambda$ for a loop of size $L$ is 
$\Lambda\sim1/L$. Consequently we expect
\begin{align}
	W \equiv b_0 &+ b_1 L\,\tr \left(m (U+U^\dagger)\right) \nonumber\\
	&+ b_2 L^2\, \tr\left(
	\partial_\mu U \partial^\mu U^\dagger\right) + \cdots
\end{align}
where $m=\mathrm{diag}(m_u,m_d,m_s)$ breaks chiral symmetry, and 
$U\equiv\exp(i\phi/F)$ with
\begin{align}
	\phi &= \phi^\dagger \nonumber \\
	&\equiv \left[
	\begin{array}{ccc}
		\pi^0/\sqrt{2}+\eta_8/\sqrt{6} & \pi^+ & K^+ \\
		\pi^- & -\pi^0/\sqrt{2}+\eta_8/\sqrt{6} & K^0 \\
		K^- & \overline{K}^0 & -2 \eta_8/\sqrt{6}
	\end{array}
	\right]
\end{align}
and $F\approx92$\,MeV.

Taking the vacuum expectation value and a logarithm, and keeping only 
the leading $\order(a)$~terms, we get
\begin{align}\label{chiral-extrap}
	\log\langle W \rangle &\approx w^{(0)} \left( 1 +
	w^{(1)}_m a\,\langle\tr \left(m (U+U^\dagger)\right)\rangle \right) 
\nonumber \\
	&\approx w^{(0)} \left( 1 + w^{(1)}_m a (2m_l + m_s) + \cdots\right).
\end{align}
Standard methods can be used to compute higher-order corrections, 
including chiral logarithms, from the expansion of $\tr \left(m 
(U+U^\dagger)\right)$, but these are
too small to be relevant to our analysis.

The leading contribution, $w^{(0)}$, is obtained from the perturbative 
analysis discussed in Section~\ref{sec:perturbation_theory}, provided 
the loops are sufficiently small to be perturbative.  We expect 
$w^{(1)}_m$ to be roughly independent of loop size since $w^{(0)}$ is 
approximately proportional to~$L/a$ (see 
Section~\ref{sec:perturbation_theory}).

We can estimate the size of $w^{(1)}_m$ from a simple argument. For 
light-quark hadrons, hadronic quantities like meson decay constants or 
baryon masses depend approximately linearly on the masses of their 
valence quarks. The mass~$m_v$ of a valence quark makes a contribution 
of  order $Q\,m_v/\Lambda$ to some hadronic quantity~$Q$, where 
$\Lambda$ is a momentum scale characteristic of the size of the hadron 
($\approx$ the chiral scale, for light-quark hadrons). From ratios of 
decay constants like $f_K/f_\pi$ or of baryon masses like 
$m(\Lambda^0)/m(p^+)$, it is clear that $m_s/\Lambda$ is of order~20\%, 
and therefore that $\Lambda\approx400$\,MeV. Empirically contributions 
from individual sea-quark masses are 3--5~times smaller than those from 
individual valence-quark masses~\cite{sea-masses}. Consequently the 
relative contribution from a sea-quark mass~$m_q$ should be roughly 
$m_q/1.2$\,GeV.

Now consider Wilson loops. The $m_q$ dependence of $\log(W_{11})$, for 
example, should be much smaller than that for a light-quark hadron 
because the loop is much smaller than the hadron. The typical radius of 
such hadrons is around~1\,fm, so we expect the relative contribution to 
$W_{11}$ from a sea-quark mass of~$m_q$ to be approximately
\begin{equation}
	\frac{a}{1\,\mathrm{fm}}\,\frac{m_q}{1.2\,\mathrm{GeV}} \approx 
\frac{am_q}{6}.
\end{equation}
Therefore we expect $w^{(1)}_m=\order(1/6)$. This implies corrections 
to our $\log(W)$s, for example, of order 1--2\% on the coarsest 
lattices and 0.3--0.5\% on the finest lattices\,---\,which is large 
compared with the statistical errors in these quantities, and therefore 
important.

In most lattice calculations we want the light-quark masses as close to 
their physical values as possible, so that lattice results reproduce 
what is seen in experiments. The situation for our Wilson loops is 
different, however. In our simulations here we are trying to isolate 
the perturbative part of the loop, in order to compare it with 
perturbation theory (not experiment), and the linear quark-mass 
dependence is a nonperturbative contamination that we want to remove. 
Consequently the precise values of the quark masses are not relevant so 
long as they are small enough that we can correct for them (or ignore 
them), which is the case here.

	\subsection{Gluon Condensate} \label{sub:gluon_condensate}
The leading gluonic nonperturbative contribution comes from the gluonic 
condensate, $\langle \alpha_s G^2/\pi\rangle$. The contribution of the 
condensate to a Wilson loop is easily calculated to leading order in 
perturbation theory:
\begin{equation}\label{eq:cond-a}
	 \delta W_\mathrm{cond} = -\frac{\pi^2}{36}\, 
\left(\frac{A}{a^2}\right)^2
		a^4 \langle \alpha_s G^2/\pi\rangle
\end{equation}
where $A$ is the loop area for planar loops. We remove this 
contribution from our Wilson loops before comparing them with 
perturbation theory. The value of the condensate is not well known, so 
we take $\langle \alpha_s G^2/\pi\rangle = 
0.0\pm0.012\,\mathrm{GeV}^4$, which covers the range of 
expectations~\cite{Ioffe:2005}. We also allow for higher-dimension 
condensate contributions by
replacing
\begin{align}\label{eq:cond-b}
	\delta W_\mathrm{cond} \to \delta W_\mathrm{cond} \big( 1
		& + w_\mathrm{cond}^{(2)} (a\Lambda_g)^2 \nonumber \\
	    &+ w_\mathrm{cond}^{(4)} (a\Lambda_g)^4 + \cdots
	\big),
\end{align}
where we take $\Lambda_g=1$\,GeV and coefficients 
$w_\mathrm{cond}^{(i)}=0\pm1$. To be certain that we do not 
underestimate errors we include 10~condensate terms in 
all~\cite{lepage:cond-correl}.

We chose the number of condensate terms here somewhat arbitrarily. Only 
results from the largest loops are affected appreciably even by the 
leading-order condensate correction, and then only by amounts of order 
a standard deviation in our final results for the coupling. While a 
leading-order condensate value of 0.006, for example, shifts $\log 
W_{23}$ by about~25\% for our largest lattice spacings, the shift is 
less than~0.1\% for the smallest lattice spacing, which is more 
important in our analysis. Smaller loops are much less sensitive: for 
example, this gluon condensate shifts $\log W_{11}$ by only~0.3\% for 
the largest lattice spacings, and by only~0.003\% for the smallest 
lattice spacings. The two Creutz ratios that involve $W_{23}$ are the 
most sensitive to condensate contributions, but even they are shifted 
by only~0.2--0.25\% for the smallest lattice 
spacings~\cite{lepage:thank-maltman}.

	\subsection{Finite-$a$ Errors} \label{sub:finite_a_errors}
In our analysis, the scale for the couplings comes from the lattice 
spacing, and the lattice spacing comes from measurements of $r_1/a$ in 
the simulations.  As for any physical quantity, lattice QCD 
measurements of $r_1$ have finite-$a$ errors; and, using an analysis 
similar to the one we outlined for Wilson loops, they should also be 
approximately linear in the sea-quark masses. Consequently we expect
\begin{equation}\label{r1-a2-dep}
	r_{1}^{\mathrm{lat}} = r_1 ( 1
	+ r^{(2)}_{1a} (a/r_1)^2  	+  r^{(1)}_{1m} r_1 (2\delta m_l+\delta 
m_s) + \cdots
	)
\end{equation}
where: $r^{(2)}_{1a}=\order(\alpha_s\approx1/3)$~\cite{running-alpha}, 
since the gluon action has no tree-level errors in~$\order(a^2)$; and 
$r^{(1)}_{1m}=\order(1/6)$, following the discussion for Wilson loops. 
Here $\delta m_q$  is the simulation's tuning error in the mass for 
sea-quark~$q$\,---\,$\delta m_l\approx m_l$ for our simulations, while 
$\delta m_s\approx0$. These corrections could affect our lattice 
spacings by as much as several percent, although the impact on 
$\alpha_0$ is suppressed by a power of $\alpha_0$ and so is much less. 
We allow for both corrections in our analysis.

\section{Analysis and Results} \label{sec:analysis_and_results}
We have 22~different short-distance quantities in our analysis, each of 
which produces a separate value for 
$\alpha_0\equiv\alphav(7.5\,\mathrm{GeV})$. These consist of $\log(W)$s 
for each of 8~Wilson loops, 6~independent Creutz ratios built from 
these loops, 7~tadpole-improved $\log(W)$s, and the tadpole improved 
bare coupling $\alpha_\mathrm{lat}/W_{11}$. We have 12~values for each 
of these quantities, with one for each configuration set in 
Table~\ref{tab:qcd-param}. In this section we discuss first the fitting 
method used for extracting~$\alpha_0$, and then we review our results.

\subsection{Constrained Fits} 	We analyze each short-distance 
quantity~$Y$ separately. We use a constrained fitting procedure, based 
upon Bayesian ideas~\cite{Lepage:2001ym}, to fit the values 
$Y_i\pm\sigma_{Y_i}$ coming from each of our configuration sets 
(Table~\ref{tab:w-mc}) to a single formula. In this procedure we 
minimize an augmented $\chi^2$~function of the form
	\begin{equation}
		\chi^2 \equiv \sum_{i=1}^{12}
		\frac{(Y_i - 
Y(a_i,(am_q)_{i},\alpha_0,y_m^{(1)},c_n,d))^2}{\sigma_{Y_i}^2}
		+ \sum_\xi \delta \chi^2_\xi
	\end{equation}
where $i$ labels the configuration set, and
	\begin{align}
		Y(a_i,&(am_q)_i,\alpha_0,y_m^{(1)},c_n,d)  \nonumber\\
		&= \left(1 + y_m^{(1)}(2am_l+am_s)_i\right)
		\sum_{n=1}^{10} c_n \alphav^n(d/a_i).
	\end{align}
The sea-quark mass dependence here is from \eq{chiral-extrap}. The 
lattice spacing in each case is determined from the simulation values 
for $(r_1/a)_i$ from each configuration set (Table~\ref{tab:qcd-param}) 
using
\begin{align}\label{a_i}
	a_i
	= \frac{r_1}{(r_1/a)_i}\,
	\left(1
	+ r^{(2)}_{1a} (a/r_1)_i^2  + r^{(1)}_{1m} (2r_1m_l)_i
	\right),
\end{align}
which follows from \eq{r1-a2-dep}, taking $\delta m_l\approx m_l$ and 
$\delta m_s\approx0$, and $r_1=0.321(5)$\,fm~\cite{Gray:2005ur}. Here 
$(r_1m_q)_i \equiv (am_q)_i (r_1/a)_i$. Given the lattice spacing, the 
coupling $\alphav(d/a)$~is computed from $\alpha_0$ by integrating 
\eq{evol-eq} numerically.

The $\chi^2$ function is minimized by varying fit parameters like 
the~$c_n$ (but not $d$ which is effectively exact). Every fit parameter 
in our procedure is constrained by an extra term or ``prior''~$\delta 
\chi^2_\xi$ in the $\chi^2$ function. The expansion parameters~$c_n$ 
from perturbation theory, for example, are constrained by
\begin{equation}
	\delta \chi^2_{c_n} = \sum_{n=1}^{10}
	\frac{(c_n-\overline{c}_n)^2}{\sigma^2_{c_n}},
\end{equation}
which implies that the fit will explore values for $c_n$ that are 
centered around $\overline{c}_n$ with a range specified by 
$\sigma_{c_n}$: $\overline{c}_n\pm\sigma_{c_n}$. For $n\le3$, we set 
$\overline{c}_n$ to the value obtained from our numerical evaluation of 
the relevant Feynman diagrams, with $\sigma_{c_n}$ equal to the 
uncertainty in that evaluation. For $n\ge4$, we set $\overline{c}_n=0$ 
and
\begin{equation}
	\sigma_{c_n} = 2.5\, \mathrm{max}(|c_1|,|c_2|,|c_3|).
\end{equation}
Thus the $c_n$s in the fit are constrained by the values obtained from 
our Feynman integrals where these are available (taking correct account 
of the uncertainties in those values), while the others are allowed to 
vary over a range that is $2.5$~times larger than the largest known 
coefficient. The factor~2.5 was chosen using the empirical Bayes 
criterion, described in~\cite{Lepage:2001ym}, applied to the 
$\log(W)$s; applying the same criterion to the other quantities would 
have given smaller factors, but we take the more conservative factor 
of~2.5 for these as well.

We include seven $c_n$s beyond the ones currently known from 
perturbation theory to illustrate an important issue. In reality there 
are infinitely many $c_n$s, but in practice the various uncertainties 
in our analysis mean that it is sensitive only to the first few. As we 
add $c_n$s the fit improves but only up to a point\,---\,$n=4$ for 
$\log(W)$s. As long as priors are included in $\chi^2$, terms can be 
added beyond this point but they have no effect on the result of the 
fit (including the error estimate) or on the quality of the fit. We add 
terms through $n=10$ to be certain we have reached this point. Our 
analysis is not sufficiently accurate to yield new information about 
$c_n$s with $n>4$ (beyond what is incorporated in the prior); but, by 
adding enough $c_n$s so that the fit results and errors cease changing, 
we guarantee that our final error estimates include the full 
uncertainty due to the fact that we have \emph{a priori} values for 
only a few of the coefficients.

Other fit parameters, like $\alpha_0$, $y_m^{(1)}$, $r_{1a}^{(2)}$, and 
$r_{1m}^{(1)}$, must also have priors:
\begin{align}
	\delta \chi^2_0 =&
	\frac{(\log(\alpha_0)-\overline{\log(\alpha_0)})^2}{\sigma_{\log(\alpha_0)}^2}
	+ \frac{(y_m^{(1)}-\overline{y}_m^{(1)})^2}{\sigma^2_{y_m^{(1)}}} 
\nonumber \\
	&+ 
\frac{(r_{1a}^{(2)}-\overline{r}_{1a}^{(2)})^2}{\sigma^2_{r_{1a}^{(2)}}}
	+ 
\frac{(r_{1m}^{(1)}-\overline{r}_{1m}^{(1)})^2}{\sigma^2_{r_{1m}^{(1)}}}
\end{align}
We constrain $\log(\alpha_0)$ to be $-1.6\pm0.5$; this prior has 
negligible effect on the fits because it is so broad (and the fits are 
so sensitive to~$\alpha_0$). Following the discussion in 
Section~\ref{sec:systematic_errors}, we set
\begin{align}
	\overline{y}_m^{(1)}  &= \overline{r}_{1m}^{(1)} = 0
	\nonumber \\
   \sigma_{y_m^{(1)}} &= \sigma_{r_{1m}^{(1)}} = 1/6.
\end{align}
We checked the width of these two priors using the empirical Bayes 
criterion and found that, in fact, this is the optimal width indicated 
by our simulation results. For $r_{1a}^{(2)}$, the empirical Bayes 
criterion suggests a width for the prior that is twice what we 
anticipated in Section~\ref{sub:finite_a_errors}:
\begin{equation}
	\overline{r}_{1a}^{(2)} = 0
	\quad\quad
	\sigma_{r_{1a}^{(2)}} = 2\alpha_s \approx 0.6.
\end{equation}
We use this more conservative prior in our fits. Higher-order 
corrections are easily added but have no impact because the corrections 
are too small to matter, given the size of our other errors.

Our simulation result for $(r_1/a)_i$, which is used to determine the 
lattice spacing $a_i$  for the $i^\mathrm{th}$ configuration set 
(\eq{a_i}), is not exact. To include its uncertainty in our analysis we 
treat $(r_1/a)_i$ as a fit parameter, to be varied while minimizing 
$\chi^2$, but with a prior whose mean is the value measured in the 
simulation and whose width is the measured uncertainty (as in 
Table~\ref{tab:qcd-param}). We can incorporate the uncertainty in the 
value of $r_1$ using the same trick, with $r_1$ as a fit parameter:
\begin{equation}
	\delta \chi^2_{r1} =
	\frac{(r_1-\overline{r}_1)^2}{\sigma_{r_1}^2} +
	\sum_{i=1}^{12}
	\frac{((r_1/a)_i - \overline{(r_1/a)}_i)^2}{\sigma^2_{(r_1/a)_i}}
\end{equation}
where $\overline{r}_1\pm\sigma_{r_1}= 
0.321\pm0.005$\,fm~\cite{Gray:2005ur}.

The~$c$ and~$b$ masses are required to convert $\alpha_0$ to 
$\alphamsb(M_Z,n_f\!=\!5)$. We account for the uncertainties in these 
masses by including them as fit parameters, with appropriate priors, 
together with fit parameters for unknown high-order terms in the $\msb$ 
$\beta$-function, and in the perturbative formulas for incorporating 
$c$~and $b$~vacuum polarization~\cite{heavy-quarks,Schroder:1998vy}. 
For the $\beta$-function, we allow for a sixth-order term 
$\beta_4\alpha_\msb^6$ in the evolution equation (analogous to 
\eq{evol-eq} for~$\alpha_V$) where $\beta_4$ is a fit parameter with a 
prior centered on $\overline{\beta}_4=0$ with width
\begin{equation}
	\sigma_{\beta_4}=\mathrm{max}(|\beta_0|,|\beta_1|,|\beta_2|,|\beta_3|)
\end{equation}
for the $\msb$~$\beta_i$s. We include analogous corrections, fit 
parameters and priors for the formulas for $c$~and $b$~vacuum 
polarization.

\subsection{Results} \label{sub:results}

\begin{figure}
	\begin{center}
	\includegraphics[scale=1.0]{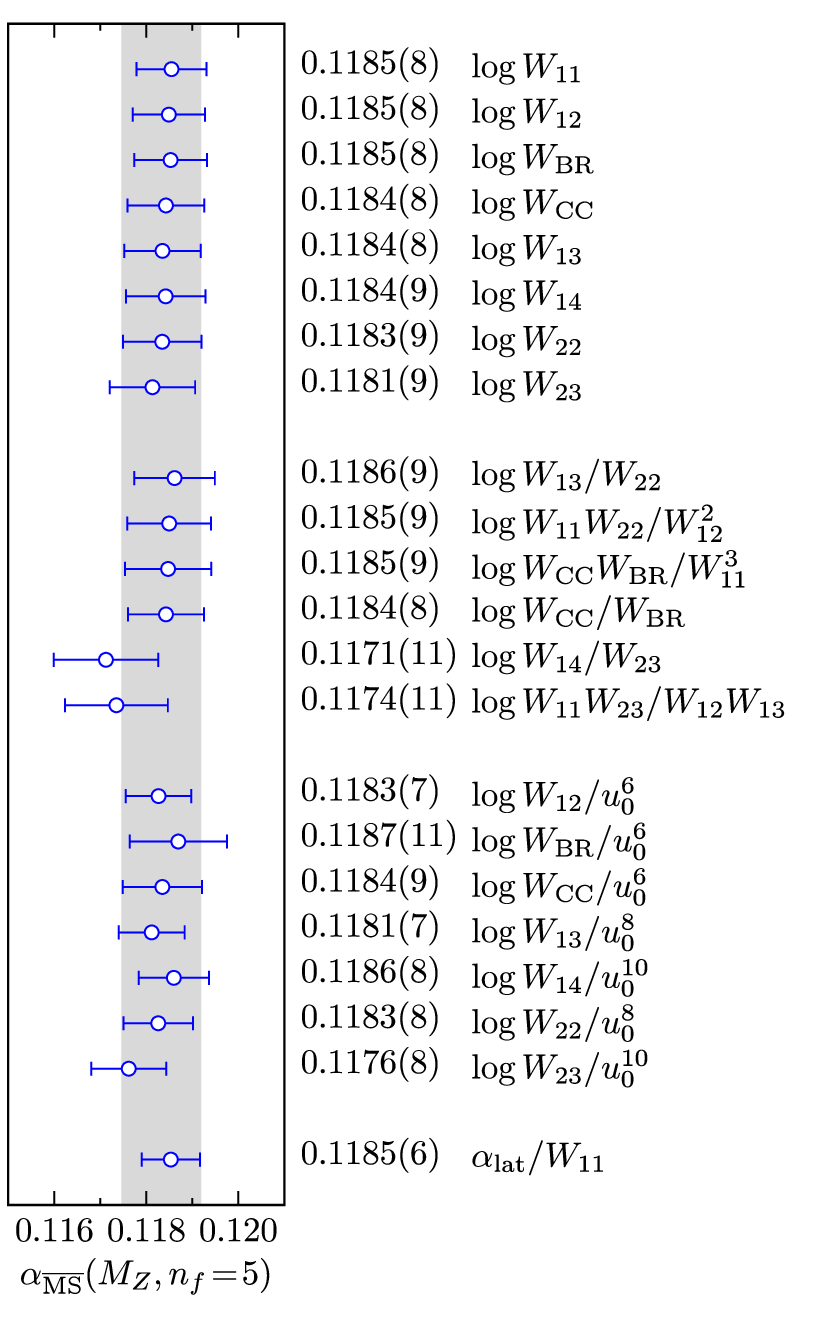}
	\end{center}
	\caption{Values for the 5-flavor $\alphamsb$ at the $Z$-meson mass 
from each of 22~short-distance quantities. The gray band indicates our 
final result, \almz. $\chi^2$ per data point is~\chialmz.}
	\label{fig:almz-results}
\end{figure}

The results from our 22~determinations of the coupling are listed and 
shown in Figure~\ref{fig:almz-results}. The gray band corresponds to 
our final result of
\begin{equation}
	\alpha_\msb(M_Z,n_f\!=\!5) = \almz
\end{equation}
which was obtained from a weighted average of all of 
22~determinations~\cite{wgtd-avg}. Our error estimate here is that of a 
typical entry in the plot; combining our results does not reduce errors 
because most of the uncertainty in each result is systematic. The 
individual results in the plot are consistent with each other: 
$\chi^2/22=\chialmz$ for the 22~entries in 
Figure~\ref{fig:almz-results}. And the fits for each quantity 
separately are excellent as well: $\chi^2/12=0.3$ to~0.6 for our fits 
to the 12~pieces of simulation data (one from each configuration set) 
for each quantity.
The results in Figure~\ref{fig:almz-results} are derived, using 
perturbation theory (Section~\ref{sec:perturbation_theory}), from the 
fit values for $\alpha_0$, which average to
\begin{equation}\label{alphav-ans}
	\alpha_0 = \alphav(7.5\,\mathrm{GeV},n_f\!=\!3) = \alv,
\end{equation}
where again the error is that of a typical result for a single 
short-distance quantity (it is \emph{not} reduced by one over the 
square root of the number of inputs).

\begin{figure}
	\begin{center}
		\includegraphics{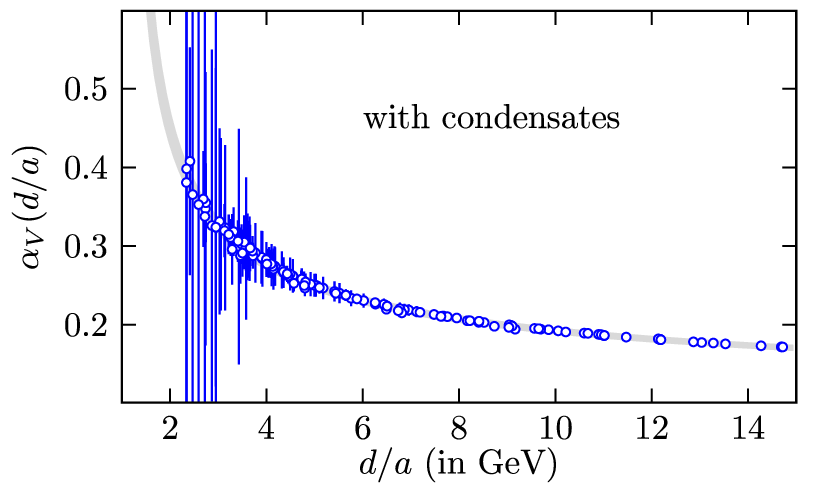}\\
		\includegraphics{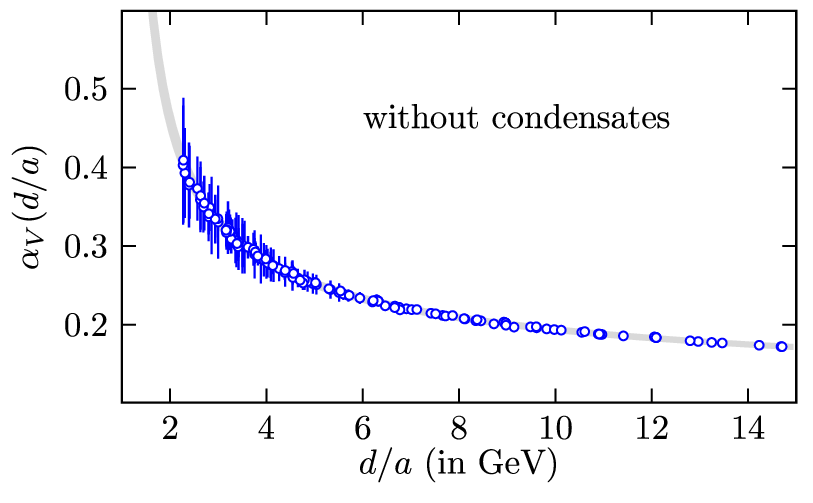}
	\end{center}
	\caption{Values for $\alphav$ versus $d/a$ from each short-distance 
quantity at each lattice spacing, with and without corrections for 
gluon condensates. The gray band shows the prediction from QCD 
evolution (\eq{evol-eq}) assuming our composite fit value 
(\eq{alphav-ans}).}
	\label{fig:alv-results}
\end{figure}

Figure~\ref{fig:alv-results} reveals more details about our fit. The 
top panel in this figure shows the values of $\alphav(d/a)$ coming from 
every short-distance quantity for every lattice spacing in our 
configuration sets. The $\alphav$s plotted here were obtained by 
refitting each piece of simulation data separately, rather than fitting 
results from all lattice spacings simultaneously as above. In these 
fits we used the values for $c_n$ with $n>3$, $w^{(1)}_m$, 
etc.\ obtained from our simultaneous fit to all lattice 
spacings~\cite{lepage:details}, which is why the individual data points 
align well with the perturbative result for $\alphav(d/a)$ (the gray 
band). The fact that different points align so well is an indication of 
the self-consistency of our perturbative analysis across all scales and 
for all quantities. The size of the error bars for different points is 
determined by the perturbative and nonperturbative uncertainties 
associated with each piece of simulation data. Points with error bars 
much larger than the uncertainties in the perturbative $\alphav$ (that 
is, much larger than the vertical width of the gray band) have little 
impact on our overall fits. The bulk of the uncertainty at low momentum 
comes from uncertainties in the gluon condensates. This is obvious when 
the results are reanalyzed without corrections for the condensates 
(bottom panel in Figure~\ref{fig:alv-results}). The most important 
simulation data is at large $d/a$, where errors are smaller than the 
plot points whether or not condensates are included.

\begin{table*}
	\begin{ruledtabular}
		\begin{tabular}{cddddddd} 			 & \multicolumn{1}{c}{$\log 
W_\mathrm{11}$} & \multicolumn{1}{c}{$\log W_\mathrm{12}$} & 
\multicolumn{1}{c}{$\log W_\mathrm{22}$} & \multicolumn{1}{c}{$\log 
W_{11}W_{22}/W_{12}^2$} & \multicolumn{1}{c}{$\log 
W_\mathrm{12}/u_0^{6}$} & \multicolumn{1}{c}{$\log 
W_\mathrm{22}/u_0^{8}$} & 
\multicolumn{1}{c}{$\alpha_\mathrm{lat}/W_{11}$}\\ \hline
                   $c_1...c_3$ &  0.1\% &  0.1\% &  0.1\% &  0.3\% &  
0.1\% &  0.1\% &  0.1\%\\
             $c_n$ for $n\ge4$ &  0.2 &  0.3 &  0.3 &  0.4 &  0.3 &  
0.4 &  0.3\\
$am_q$, $r_1m_q$ extrapolation &  0.1 &  0.1 &  0.0 &  0.1 &  0.1 &  
0.1 &  0.0\\
     $(a/r_1)^2$ extrapolation &  0.2 &  0.3 &  0.4 &  0.3 &  0.2 &  
0.2 &  0.0\\
            $(r_1/a)_i$ errors &  0.4 &  0.4 &  0.4 &  0.3 &  0.3 &  
0.3 &  0.3\\
                  $r_1$ errors &  0.3 &  0.3 &  0.3 &  0.3 &  0.3 &  
0.3 &  0.3\\
              gluon condensate &  0.1 &  0.1 &  0.1 &  0.2 &  0.1 &  
0.1 &  0.1\\
            statistical errors &  0.0 &  0.0 &  0.0 &  0.1 &  0.0 &  
0.0 &  0.0\\
             $V\to\msb\to M_Z$ &  0.1 &  0.1 &  0.1 &  0.1 &  0.1 &  
0.1 &  0.1\\
\hline
               Total  &  0.6\% &  0.6\% &  0.7\% &  0.7\% &  0.6\% &  
0.6\% &  0.5\%

		\end{tabular}
	\end{ruledtabular}
	\caption{Sources of uncertainties in determinations of 
$\alphamsb(M_Z,n_f\!=\!5)$ from various short-distance quantities. 
Uncertainties are given as percentages of the final result in each 
case.}
	\label{tab:almz-errors}
\end{table*}

It is useful to separate our error estimates into component pieces. The 
error estimate produced by our fitting code for a quantity like 
$\alphamsb$ is approximately linear in all the variances~$\sigma^2$ 
that appear in the $\chi^2$ function:
\begin{align}\label{lin-approx}
	\sigma_{\alphamsb}^2 \approx&
	\sum_{i=1}^{12} \sigterm{Y_i} + \sum_{n=1}^{10} \sigterm{c_n}
	+ \sigterm{y^{(1)}_m} \nonumber \\
	 &+ \sigterm{r^{(1)}_{1m}} + \sigterm{r^{(2)}_{1a}} + \cdots
\end{align}
This works when errors are small, as they are here. To isolate the part 
of the total error that is associated with the statistical 
uncertainties in the $Y_i$, for example, the fit is rerun but with the 
corresponding variances rescaled by a factor~$f$ close to one 
($f=1.01$, for example):
\begin{equation}
	\sigma^2_{Y_i}\to f \sigma^2_{Y_i}
\end{equation}
for $i=1\ldots12$. Then
\begin{equation}\label{diff-eq}
	\frac{\sigma^2_{\alphamsb}(f)-\sigma^2_{\alphamsb}(f\!=\!1)}{f-1}
	\approx \sum_{i=1}^{12} \sigterm{Y_i}
\end{equation}
The square root of this quantity is the part of the total error due to 
the statistical uncertainties in the~$Y_i$. This procedure can be 
repeated for each prior or group of priors that contributes to the 
$\chi^2$ function. The sum of the variances obtained in this way for 
each part of the total error should equal $\sigma^2_{\alphamsb}$; if it 
does not, errors may not be sufficiently small to justify the linear 
approximation in~\eq{lin-approx}~\cite{pathology}.

In Table~\ref{tab:almz-errors} we present error budgets computed in 
this fashion for a sample of our determinations of $\alphamsb(M_Z)$. 
This table shows that our largest errors come from uncertainties in the 
perturbative coefficients with~$n\ge4$, statistical errors in the 
simulation values for $(r_1/a)_i$, systematic uncertainties in the 
physical value for~$r_1$, and finite-$a$ lattice errors in $r_1$. 
Uncertainties in the parameters used to convert 
$\alpha_0=\alphav(7.5\,\mathrm{GeV},n_f\!=\!3)$ into 
$\alphamsb(M_Z,n_f\!=\!5)$ have negligible impact. Also negligible are 
uncertainties due to the gluon condensate, and statistical errors in 
the Wilson loops.

Our errors are greatly reduced because we can bound the size of 
perturbative coefficients $c_n$ for $n=4$ and beyond. This is possible 
because we are fitting simulation data from six different lattice 
spacings simultaneously. As noted in~\cite{Mason:2005zx}, the $n=4$ 
coefficients are large, particularly for $\log(W)$s where typically our 
fits imply $c_4/c_1\approx -4(2)$. As expected, perturbative 
higher-order coefficients are smaller for other quantities: for 
example, we find typically $c_4/c_1\approx -2(2)$ for tadpole-improved 
loops. The fit results for $c_4/c_1$ and $c_5/c_1$ for each of our 
short-distance quantities are given in Table~\ref{tab:pth-coef}.

We tested the stability of our analysis procedure in several ways:
\begin{itemize}
	\item \emph{Discarding simulation data:} Dropping data for any one of 
the lattice spacings gives results that are almost identical to our 
final result: the value of $\alphamsb(M_Z)$ varies by no more 
than~0.12\% from our final result, and its uncertainty ranges 
between~0.00083 and~0.00093. Dropping the two smallest lattice 
spacings, which are the most important, shifts $\alphamsb(M_Z)$ 
to~0.1176(14). Keeping just the four, three and two smallest lattice 
spacings gives~0.1183(9), 0.1180(10), and~0.1179(10), respectively (for 
sets~4--12, 7--12, and 9--12).

	\item \emph{Perturbation theory scale changes:} Our results do not 
depend strongly on the choice of scale~$d/a$ used in the perturbation 
theory for each quantity. Re-expanding our perturbation theory for 
$d\to d/1.5$ or $d\to 1.5 d$, for example, shifts the overall 
$\alphamsb(M_Z)$ to~0.1181(8) or~0.1184(8), 
respectively~\cite{Maltman:2008bx}.

	\item \emph{$\msb$ throughout:} Re-expressing the perturbation theory 
for each quantity in terms of $\alphamsb$ in place of $\alphav$ gives 
almost the same overall results, 0.1185(10), but leads to significantly 
larger high-order coefficients in perturbation theory (2.5~times larger 
for small loops), somewhat greater dispersion between results from 
different quantities ($\chi^2/22$ of~0.5 instead of~0.2), and larger 
uncertainties in the results from most quantities. The scale-setting 
procedure used to select the~$d$s is tailored specifically for 
$\alphav$~expansions; this is reflected by these results.

	\item \emph{Adding more/fewer perturbative terms:} We allow terms up 
through tenth order in the perturbative expansions for the various 
short-distance quantities. Adding further terms has no impact on our 
results. Restricting perturbation theory to only fourth or fifth order 
also leaves our final result unchanged. Fitting is impossible with 
fewer than four terms: with three terms fits for individual Wilson 
loops, for example, to data from all 12~configuration sets are poor, 
with $\chi^2/12$ becoming as large as~1.9 (rather than~0.4); and the 
couplings coming from the 22~different short-distance quantities 
disagree with each other, giving $\chi^2/22=1.45$ (rather than~0.16).

	\item \emph{Adding more/fewer nonperturbative terms:} Adding 
higher-order terms in the chiral expansion in sea-quark masses 
(\eq{chiral-extrap}) or further terms in the gluon-condensate expansion 
(\eq{eq:cond-b}) does not change our final result at all. Omitting all 
corrections for the gluon condensates increases $\alphamsb(M_Z)$ by two 
thirds of a standard deviation, to~0.1189(7). If we keep only the three 
smallest lattice spacings, which are the least sensitive to 
nonperturbative effects, we get~0.1180(10) whether or not the gluon 
condensates are included. We cannot fit all of our simulation data if 
we omit the chiral correction. Fitting without chiral corrections 
becomes possible if we keep only the subset of our data with 
$m_{u/d}/m_s\approx0.2$ (sets~2, 5, 7, 10, and~12); this gives 
$\alphamsb(M_Z)=0.1181(9)$. (Our fit to $\log(W_{11})$ gives
\begin{equation}
	w_m^{(1)} = -0.18\,(6)\quad\quad r_{1m}^{(1)} = -0.08\,(8),
\end{equation}
which is typical of the other fits.)

\end{itemize}
Each of the variations examined here gives results that agree with our 
final result to within a standard deviation, suggesting that we have 
not underestimated the uncertainty in our result.

Our new result is one standard deviation above our previous result from 
Wilson loops~\cite{Mason:2005zx}, $\alphamsb(M_Z)=0.1170(12)$, and has 
an error that is 33\%~smaller. Our new analysis differs in two 
important ways from our earlier work. First we include more lattice 
spacings, including one that is 50\%~smaller than the smallest we used 
before. (We used only configuration sets~1, 5 and~7 before.) This 
significantly reduces the errors. Second we now use more accurate 
values for $r_1/a$. These reduce uncertainties in the ratios of lattice 
spacings from different configuration sets, to a third of what they 
were in our earlier analysis. This matters since comparing results at 
different lattice spacings bounds the uncalculated high-order 
perturbation theory coefficients in our analysis ($c_n$ for $n\ge4$). 
We are also allowing for larger finite-$a$ errors in $r_1/a$ on the 
coarsest lattices than we did previously. The changes in $r_1/a$, 
together with the smaller lattice spacing, account for most of the 
increase in our final result.

Another change, which has less impact, is the inclusion of possible 
higher-dimension condensates. We also now do a more systematic analysis 
of effects due to the sea-quark mass, fitting results with many 
different masses, but the effect on our final result is small. Finally, 
we now use better scales~$d/a$ for the Creutz ratios and 
tadpole-improved loops than in our previous 
analysis~\cite{Hornbostel:2002af}. Using the new scales shifts our 
final result up by only a third of a standard deviation, but the 
dispersion between results from different short-distance quantities is 
decreased from $\chi^2/22=0.6$ to~\chialmz.

\section{Conclusions} \label{sec:conclusions}
Any high-precision determination of $\alpha_s$ based upon lattice QCD 
simulations has to address several key issues:
\begin{itemize}
	\item \emph{Finite-Lattice-Spacing Errors:} Errors due to the finite 
lattice-spacing can enter in two ways. First they affect lattice 
determinations of the physical quantity or quantities used to set the 
scale of the coupling. In our analysis we use simulation values for 
$r_1/a$, from the static-quark potential, to determine ratios of scales 
from different configuration sets, and simulation values for the 
$\Upsilon^\prime$--$\Upsilon$ mass difference to set the overall 
scale~\cite{Gray:2005ur}. In each case we use data from multiple 
lattice spacings to bound finite-$a$ errors, which are small because we 
use highly-improved discretizations in our simulations. The second 
source of finite-$a$ errors, for some analyses (but not ours), is the 
lattice determination of the short-distance quantity that is compared 
with perturbation theory (to extract~$\alpha_s$). A short-distance 
quantity that is defined in continuum QCD\,---\,for example, changes 
$V(r_a)-V(r_b)$ in the static-quark potential for small 
$r$s~\cite{Mason:2005zx,loop-paper}, or current-current correlators for 
$c$-quark currents~\cite{Allison:2008xk}\,---\,will have finite-$a$ 
errors that must be included in the final error analysis. The use of 
multiple lattice spacings is again important. This is not an issue for 
us here because we analyze our short-distance quantities using lattice 
QCD perturbation theory, which treats finite-$a$ effects exactly (that 
is, to all orders in~$a$, order-by-order in $\alphav$). Both the 
simulation results and the perturbation theory for our 
22~short-distance quantities are free of finite-$a$ errors. This 
greatly facilitates our use of results from multiple lattice spacings 
to bound uncalculated higher-order terms from perturbation theory.

	\item \emph{Truncation Errors from Perturbation Theory:} The coupling 
is determined by comparing perturbation theory with (nonperturbative) 
simulation results for a short-distance quantity. Generally the 
perturbation theory is known through only a few low orders 
in~$\alpha_s$. The error analysis for any determination of the coupling 
must account for the uncalculated (but certainly present) terms from 
higher-order perturbation theory. We not only account for the 
possibility of higher-order terms (through tenth order), using our 
Bayesian priors, but also attempt to estimate the size of these 
corrections by comparing values of our short-distance quantities at 
five different momentum scales~$d/a$, corresponding to our five lattice 
spacings. We find sizable contributions from high-order terms, 
particularly for $\log(W)$s: leaving them out would shift our final 
result for the coupling down by one to two standard deviations (and 
lead to poor fits for most of our short-distance quantities). The 
agreement between our 22~different short-distance quantities, some with 
very different perturbative expansions (see 
Section~\ref{sec:perturbation_theory}), is important evidence that we 
have analyzed truncation errors correctly.

	\item \emph{Sea-Quark Vacuum Polarization:} In our previous 
analysis~\cite{Mason:2005zx}, we showed that the coupling is quite 
sensitive to contributions from the vacuum polarization of sea quarks: 
$\alphamsb(M_Z)$ is 30\% smaller when all quark vacuum polarization is 
omitted. It is therefore important to include vacuum polarization from 
all three light quarks. Vacuum polarization corrections from heavy 
quarks ($c$, $b$ and $t$) can be computed using perturbation theory, 
but light quarks ($u$, $d$ and $s$) can only be incorporated 
nonperturbatively. In the past we have used simulations with fewer than 
three light-quarks and extrapolated to $n_f=3$ ($1/\alphamsb(M_Z)$ 
appears to be reasonably linear in $n_f$)~\cite{Davies:1997mg}. Here 
(and in our earlier paper~\cite{Mason:2005zx}) contributions from all 
three light-quarks are included in the configurations provided to us by 
the MILC collaboration. We also account for the small but (barely) 
measurable dependence upon the sea-quark masses.

	\item \emph{Other Lattice and Nonperturbative Artifacts:} Usually one 
must worry about the finite volume of the lattice in a QCD simulation. 
Our Wilson loops, however, are about as ultraviolet singular as is 
possible on a lattice, and so are completely insensitive to the volumes 
of our lattices (2.5\,fm across). Another issue, for continuum as well 
as lattice determinations of the coupling, is the possibility of 
nonperturbative contributions to the short-distance quantity. Our 
quantities are sufficiently short-distance that we do not expect 
appreciable nonperturbative contamination. We nevertheless allowed for 
nonperturbative contributions from both gluons and quarks. The expected 
size of nonperturbative contributions varies widely over our set of 
22~different short-distance quantities and 6~different lattice 
spacings. The excellent agreement among all of our results is strong 
evidence that we understand these systematic errors.
\end{itemize}

In this (and our previous) paper, we have addressed all of these 
issues. We have extended our earlier analysis of the strong coupling 
constant from Wilson loops in lattice QCD (and hadronic spectroscopy) 
to include results from 22~different short-distance quantities computed 
on 12~different lattices, with 6~distinct lattice spacings and a 
variety of sea-quark masses. We extracted a new value for the QCD 
coupling by comparing these $22\times12= 264$ different pieces of 
simulation data, varying by a factor of seven in momentum scales ($d/a$ 
from~2.1 to~14.7\,GeV), with perturbation theory. Our result, 
$\alphamsb(M_Z,n_f\!=\!5)=\almz$, is in excellent agreement with our 
previous result from Wilson loops~\cite{Mason:2005zx}, 0.1170\,(12), 
and also with non-lattice determinations: for example, the world 
averages 0.1176\,(20) from~\cite{pdg} and 0.1189\,(10) 
from~\cite{Bethke:2006ac}. Our new result also agrees well with our 
very recent result, 0.1174\,(12), from current-current correlators 
computed using lattice QCD~\cite{Allison:2008xk}.

While they are derived from the Wilson loops, our Creutz ratios and 
tadpole-improved loops provide coupling-constant information that is 
independent from that coming from the loops directly. This is because 
the highly ultraviolet contributions that dominate the loops largely 
cancel in the other quantities, making the latter more infrared. 
Consequently both perturbative and nonperturbative behavior differs 
significantly from quantity to quantity. This is particularly true of 
the sensitivity to nonperturbative contributions: for example, our most 
infrared Creutz ratios are more than 100 times more sensitive to gluon 
condensates than our most ultraviolet loops. That all of our quantities 
agree on the coupling (Figure~\ref{fig:almz-results}) is strong 
evidence that we understand the systematic errors involved.

The close agreement of our results with non-lattice determinations of 
the coupling is a compelling quantitative demonstration that the 
perturbative QCD of jets, and the QCD of lattice simulations, which 
encompass both perturbative and nonperturbative phenomena, are the same 
theory. It is also further evidence that the simulation methods we use 
are valid. While early concerns about the light-quark discretization 
used here have been largely 
addressed~\cite{Sharpe:2006re,Bernard:2007eh}, it remains important to 
test the simulation technology of lattice QCD at increasing levels of 
precision given the critical importance of lattice results for 
phenomenology~\cite{rooting}.

We thank K.\ Maltman for discussions and comments. This work was 
supported by NSERC (Canada), STFC (UK), the NSF and DOE (USA), the 
Leverhulme Trust, and by SciDAC/US Lattice QCD Collaboration 
allocations at Fermilab. Results for the smallest lattice spacing 
required resources from the Argonne Leadership Computing Facility at 
Argonne National Laboratory, which is supported by the Office of 
Science of the U.S. Department of Energy under contract 
DE-AC02-06CH11357. We also acknowledge the use of Chroma for part of 
our analysis~\cite{Edwards:2004sx}; and we thank Steve Gottlieb, Doug 
Toussaint and their collaborators in the MILC collaboration for the use 
of their gluon configurations and for unpublished values of~$r_1/a$.

\bibliographystyle{plain}

\begin{thebibliography}{16}
\bibitem{Mason:2005zx}
  Q.~Mason {\it et al.}  [HPQCD Collaboration],
    Phys.\ Rev.\ Lett.\  {\bf 95}, 052002 (2005)
  [arXiv:hep-lat/0503005].

\bibitem{Davies:2003ik}
  C.~T.~H.~Davies {\it et al.}  [HPQCD Collaboration],
    Phys.\ Rev.\ Lett.\  {\bf 92}, 022001 (2004)
  [arXiv:hep-lat/0304004].


\bibitem{Lepage:1992xa}
  G.~P.~Lepage and P.~B.~Mackenzie,
    Phys.\ Rev.\  D {\bf 48}, 2250 (1993)
  [arXiv:hep-lat/9209022].

\bibitem{Brodsky:1982gc}
  S.~J.~Brodsky, G.~P.~Lepage and P.~B.~Mackenzie,
      Phys.\ Rev.\  D {\bf 28}, 228 (1983).

\bibitem{heavy-quarks} We incorporate the $c$ and $b$ quarks using 
formulas from K. G. Chetyrkin, B. A. Kniehl and M. Steinhauser, Phys. 
Rev. Lett. 79, 2184 (1997) with quark masses of 
$m_c(m_c)=1.268(9)$\,GeV
from Allison et al. [arXiv:0805.2999] (2008), and 
$m_b(m_b)=4.18(4)$\,GeV from the HPQCD Collaboration, in preparation 
(2008). For masses, see also
J.~H.~K\"{u}hn, M.~Steinhauser and C.~Sturm,
Nucl.\ Phys.\  B {\bf 778}, 192 (2007)
[arXiv:hep-ph/0702103].

\bibitem{Schroder:1998vy} Coupling $\alphav(\mu)$ is defined in terms 
of $\alphamsb(\overline\mu)$, where 
$\overline\mu\equiv\mu\exp(-5/6)$~\cite{Lepage:1992xa,Brodsky:1982gc}, 
using the static-quark potential. The potential has been computed 
through third order:
  Y.~Schroder,
    Phys.\ Lett.\  B {\bf 447}, 321 (1999)
  [arXiv:hep-ph/9812205].
  Its expansion defines $\alphav$ through order~$\alphamsb^3$. Then 
$\beta$-function coefficients $\beta_0$ through $\beta_3$ for 
$\alpha_V$ can be computed from the $\beta$-function coefficients for 
$\alpha_\msb$, which are given in T.~van Ritbergen, J.~A.~M.~Vermaseren 
and S.~A.~Larin,  Phys.\ Lett.\ {\bf B400}, 379 (1997).

\bibitem{no-V} We are not yet able to update the analysis in our 
previous paper of the static-quark potential.

\bibitem{loop-paper} Q.\ Mason and H.\ Trottier, in preparation (2005); 
Q.\ Mason, Cornell University Ph.D.\ thesis (2004).


\bibitem{Hornbostel:2002af}
The lowest-order procedure (from~\cite{Lepage:1992xa}) for setting 
scales sometimes gives scales that are misleadingly small. A simple 
test for this problem, together with a more robust procedure for 
setting scales, is presented in:
  K.~Hornbostel, G.~P.~Lepage and C.~Morningstar,
    Phys.\ Rev.\  D {\bf 67}, 034023 (2003)
  [arXiv:hep-ph/0208224].
  The scales for most of the tadpole-improved loops (all but $W_{13}$ 
and $W_{14}$) and all of the Creutz ratios do not pass this test. For 
these cases, we have used the more robust procedure to obtain new 
scales, which we list in Table~\ref{tab:pth-coef}. Consequently scales 
for these quantities differ from what we used in~\cite{Mason:2005zx}. 
The resulting changes in our final results are not large since we are 
working to very high orders (and therefore scale changes have a small 
impact, as illustrated in Section~\ref{sub:results}).



\bibitem{lepage:erratum} In footnote~[8]  of \cite{Mason:2005zx} we 
erroneously said that the expansion for $\alpha_V$ in terms of 
$\alphamsb$ has no terms past third order. There and here, it is the 
beta function (\eq{evol-eq}) that we define to have no terms 
beyond~$\beta_3$. We thank K.\ Maltman for pointing out this 
inconsistency.




\bibitem{MILC} C. Aubin et al, Phys. Rev. D70:094505, 2004; and MILC 
Collaboration, private communication.

\bibitem{Gray:2005ur}
  A.~Gray, I.~Allison, C.~T.~H.~Davies, E.~Dalgic, G.~P.~Lepage, 
J.~Shigemitsu and M.~Wingate [HPQCD Collaboration],
    Phys.\ Rev.\  D {\bf 72}, 094507 (2005)
  [arXiv:hep-lat/0507013].
  The value for $r_1$ given in this paper has been extensively checked 
in many other calculations, some involving light quarks and others 
heavy quarks: see, for example,~\cite{Davies:2003ik}.

\bibitem{Aubin:2004wf}
  C.~Aubin {\it et al.},
      Phys.\ Rev.\  D {\bf 70}, 094505 (2004)
  [arXiv:hep-lat/0402030].

\bibitem{sea-masses}
This is quite obvious from lattice QCD simulations. It is perhaps 
because sea-quark contributions are suppressed by $1/N_c$, where 
$N_c=3$ is the number of QCD colors, in a $1/N_c$~expansion of QCD.

\bibitem{Ioffe:2005}
For a recent analysis of condensate values see B.\ L.\ Ioffe, Prog. 
Part. Nucl.
Phys. 56, 232 (2006) [arXiv:hep-ph/0502148].

\bibitem{lepage:cond-correl}
The corrections that remove the gluon condensates from the Monte Carlo 
results for a Wilson loop introduce correlations into the covariance 
matrix for results from different lattice spacings. This is because the 
corrections for different lattice spacings are related to each other 
(by Eqs.\~(\ref{eq:cond-a}) and~(\ref{eq:cond-b}), where the condensate 
values are same for different lattice spacings). It is important to 
include this correlation when fitting the dependence upon lattice 
spacing.

\bibitem{lepage:thank-maltman} We thank K.\ Maltman for an extensive 
discussion about the role of condensates here.

\bibitem{running-alpha} Rather than approximating $\alphav$ by a 
constant, we could put in the running coupling at a scale like $1/a$. 
In this case, however, it makes little difference to the final fits. 
Our analysis is not sufficiently accurate to resolve the difference 
between a constant and a running coupling in the $a^2$ corrections.

\bibitem{Lepage:2001ym}
  G.~P.~Lepage, B.~Clark, C.~T.~H.~Davies, K.~Hornbostel, 
P.~B.~Mackenzie, C.~Morningstar and H.~Trottier,
    Nucl.\ Phys.\ Proc.\ Suppl.\  {\bf 106}, 12 (2002)
  [arXiv:hep-lat/0110175].

\bibitem{wgtd-avg} We average our 22 different results weighted by 
their inverse variances, giving more weight to results with smaller 
variances. The variance for our composite result is the inverse of the 
average of the inverse variances from the separate determinations.

\bibitem{lepage:details} Specifically we fit each single piece of 
simulation data separately using the fitting function described in 
Sections~\ref{sec:systematic_errors} and~\ref{sec:analysis_and_results} 
but with priors taken from the means and standard deviations obtained 
for each parameter other than $\alpha_0$ in our overall fits for each 
short-distance quantity at all lattice spacings. Fit 
parameter~$\alpha_0$ is effectively unconstrained by the prior in these 
individual fits since we use the same prior as in the overall fits (to 
multiple lattice spacings). Parameter $\alpha_0$ is the main parameter 
adjusted by the individual fits.

\bibitem{pathology} Occasionally the difference in \eq{diff-eq} comes 
out negative. This can be caused by instabilities in the fit, in which 
case a different choice of~$f$ might fix the problem. It can also be 
the case that a term enters \eq{lin-approx} with a negative sign. In 
such cases we use the absolute value of \eq{diff-eq} for the partial 
variance.


\bibitem{Maltman:2008bx}
  The following paper, which was posted after the initial version of 
our paper, reanalyzes some of
  our Monte Carlo data using a different reorganization of perturbation 
theory.
  Results from that paper agree to within a standard deviation with our 
results.
  K.~Maltman, D.~Leinweber, P.~Moran and A.~Sternbeck,
    arXiv:0807.2020 [hep-lat].

\bibitem{Allison:2008xk}
  I.~Allison {\it et al.},
      arXiv:0805.2999 [hep-lat].

\bibitem{Davies:1997mg}
  C.~T.~H.~Davies, K.~Hornbostel, G.~P.~Lepage, P.~McCallum, 
J.~Shigemitsu and J.~H.~Sloan,
    Phys.\ Rev.\  D {\bf 56}, 2755 (1997)
  [arXiv:hep-lat/9703010].
  This paper found $\alphamsb(M_Z)=0.1174\,(24)$ by extrapolating from 
data with $n_f=0$ and $n_f=2$.


\bibitem{pdg} A world average of 0.1176(20) is given in W.-M. Yao {\it 
et al.} (Particle Data Group), J.\ Phys.\ {\bf G 33}, 1 (2006).

\bibitem{Bethke:2006ac}
  A world average of 0.1189(10) is given in S.~Bethke,
    Prog.\ Part.\ Nucl.\ Phys.\  {\bf 58} (2007) 351
  [arXiv:hep-ex/0606035].

\bibitem{Sharpe:2006re}
  S.~R.~Sharpe,
    PoS {\bf LAT2006}, 022 (2006)
  [arXiv:hep-lat/0610094].

\bibitem{Bernard:2007eh}
  C.~Bernard, M.~Golterman, Y.~Shamir and S.~R.~Sharpe,
    Phys.\ Rev.\  D {\bf 77}, 114504 (2008)
  [arXiv:0711.0696 [hep-lat]].

\bibitem{rooting} The main formal concern about staggered quarks lies 
in the use of the fourth-root of the staggered-fermion determinant to 
remove redundant ``tastes'' from the theory. Errors connected with this 
approximation vanish with the lattice spacing, and therefore 
uncertainties in our final results due to this approximation are 
included in our estimates of the uncertainties due to the~$am_q$ and 
$(a/r_1)^2$~extrapolations (Table~\ref{tab:almz-errors}).

\bibitem{Edwards:2004sx}
  R.~G.~Edwards and B.~Joo  [SciDAC Collaboration and LHPC 
Collaboration and
                  UKQCD Collaboration],
    Nucl.\ Phys.\ Proc.\ Suppl.\  {\bf 140}, 832 (2005)
  [arXiv:hep-lat/0409003].

\end{thebibliography}

\end{document}